\documentclass{aastex631}
\shorttitle{Unusual Composition of $\rm^3$He-Rich Solar Energetic Particles}
\shortauthors{Bu\v{c}\'ik et al.}
\graphicspath{{./}{figures/}}

\begin{document}

\title{Origin of Unusual Composition of $^3$He-Rich Solar Energetic Particles}

\correspondingauthor{Radoslav  Bu\v{c}\'ik}
\email{radoslav.bucik@swri.org}

\author[0000-0001-7381-6949]{Radoslav  Bu\v{c}\'ik}
\affiliation{Southwest Research Institute, San Antonio, TX 78238, USA}

\author[0000-0003-2169-9618]{Glenn M. Mason}
\affiliation{Applied Physics Laboratory, Johns Hopkins University, Laurel, MD 20723, USA}


\author[0000-0002-9242-2643]{Sargam M. Mulay}
\affiliation{School of Physics and Astronomy, University of Glasgow, Glasgow, G12 8QQ, UK}

\author[0000-0003-1093-2066]{George C. Ho}
\affiliation{Southwest Research Institute, San Antonio, TX 78238, USA}

\author[0000-0002-7388-173X]{Robert F. Wimmer-Schweingruber}
\affiliation{Institut f\"{u}r Experimentelle und Angewandte Physik, Christian-Albrechts-Universit\"{a}t zu Kiel, 24118 Kiel, Germany}

\author[0000-0002-4240-1115]{Javier Rodr\'iguez-Pacheco}
\affiliation{Universidad de Alcal\'a, Space Research Group, 28805 Alcal\'a de Henares, Spain}



\begin{abstract}

We examine $^3$He-rich solar energetic particles (SEPs) detected on 2023 October 24--25 by Solar Orbiter at 0.47\,au. The measurements revealed that heavy-ion enhancements increase irregularly with mass, peaking at S. C, and especially N, Si, and S, stand out in the enhancement pattern with large abundances. Except for $^3$He, heavy ion spectra can only be measured below 0.5\,MeV\,nucleon$^{-1}$. At 0.386\,MeV\,nucleon$^{-1}$, the event showed a huge $^3$He/$^4$He ratio of 75.2$\pm$33.9, larger than previously observed. Solar Dynamics Observatory extreme ultraviolet data showed a mini filament eruption at the solar source of $^3$He-rich SEPs that triggered a straight tiny jet. Located at the boundary of a low-latitude coronal hole, the jet base is a bright, small-scale region with a supergranulation scale size. The emission measure provides relatively cold source temperatures of 1.5 to 1.7\,MK between the filament eruption and nonthermal type III radio burst onset. The analysis suggests that the emission measure distribution of temperature in the solar source could be a factor that affects the preferential selection of heavy ions for heating or acceleration, thus shaping the observed enhancement pattern. Including previously reported similar events indicates that the eruption of the mini filament is a common feature of events with heavy-ion enhancement not ordered by mass. Surprisingly, sources with weak magnetic fields showed extreme $^3$He enrichment in these events. Moreover, the energy attained by heavy ions seems to be influenced by the size and form of jets.

\end{abstract}



\section{Introduction} \label{sec:intro}

Owing to their low intensities and the small size of the solar source, $^3$He-rich solar energetic particles (SEPs) are one of the least explored energetic particle populations in the heliosphere. They show enormous enrichment of the rare $^3$He isotope by factors up to 10$^4$ above the coronal or solar wind abundance \citep{2007SSRv..130..231M,2021LNP...978.....R,2021FrASS...8..164R}. Heavy (Ne--Fe) and ultra-heavy ions (mass $>$70\,AMU) are enhanced by a factor of 3--10 and $>$100, respectively, independent of the amount of $^3$He enhancement \citep[e.g.,][]{1986ApJ...303..849M,1994ApJS...90..649R,2022ApJS..263...22H}. The abundance enhancement of heavy ions monotonically increases (decreases) with ion mass (charge-to-mass ratio $Q$/$M$) \citep{2004ApJ...606..555M,2004ApJ...610..510R}.

Solar sources of $^3$He-rich SEPs have been associated with extreme-ultraviolet (EUV) jets \citep[e.g.,][and references therein]{2020SSRv..216...24B}, indicating acceleration at magnetic reconnection sites involving field lines open to interplanetary space \citep{2002ApJ...571L..63R,2006ApJ...639..495W}. It has been suggested that anomalous abundances of $^3$He-rich SEPs are the signature of unique acceleration or heating mechanisms. Most models involve resonant interaction with plasma waves \citep[e.g.,][]{1978ApJ...224.1048F,1992ApJ...391L.105T,1997ApJ...477..940R,2004ApJ...613L..81L,2006ApJ...636..462L,2020ApJ...888...48K}. Other mechanisms encompass ion pickup in reconnection exhaust \citep{2009ApJ...700L..16D}, Coulomb energy losses \citep{2018ApJ...862....7M}, or the first ionization potential (FIP) process \citep{2023ApJ...951...86L}.

In this paper, we investigate the 2023 October 24 $^3$He-rich SEP event observed by Solar Orbiter at 0.47\,au, showing extremely high $^3$He abundance with an unusual pattern of heavy-ion elemental composition and species with mass $\ge$4\,AMU measured only at low energies ($<$0.5\,MeV\,nucleon$^{-1}$). We include similar, previously reported events \citep{2002ApJ...565L..51M,2016ApJ...823..138M,2023ApJ...957..112M,2023AA...673L...5B} and discuss the properties of solar sources given the observed enrichment anomalies. With a novel approach, we estimate the heavy ion abundance pattern in the 2023 October 24 event using the emission measure temperature distribution at the solar source for models by \citet{1978ApJ...224.1048F} and \citet{1997ApJ...477..940R}. 

\section{Observations -- 2023 October 24 event} \label{sec:obser}

\subsection{Instruments} \label{subsec:instr}

The 2023 October 24 event elemental composition is obtained by the Suprathermal Ion Spectrograph (SIS) of the Energetic Particle Detector (EPD) suite \citep{2020AA...642A...7R} aboard Solar Orbiter \citep{2020AA...642A...1M}. SIS is a time-of-flight mass spectrometer that measures elemental composition from H through ultra-heavy nuclei in the energy range of 0.1--10\,MeV\,nucleon$^{-1}$. SIS has two telescopes, one (SIS-a) pointing at 30$^{\circ}$ (sunward) and the other (SIS-b) at 160$^{\circ}$ (anti-sunward) to the west of the spacecraft-Sun line. We also make use of solar wind plasma and interplanetary magnetic field (IMF) measurements from Solar Wind Plasma Analyzer \citep[SWA,][]{2020AA...642A..16O} and the magnetometer MAG \citep{2020AA...642A...9H} on Solar Orbiter, respectively. The solar source characteristics of SEPs were explored using the near-Earth Solar Dynamics Observatory (SDO) with 0.6\,arcsec\,pixel$^{-1}$ and 12\,s cadence Atmospheric Imaging Assembly \citep[AIA,][]{2012SoPh..275...17L} and 0.5\,arcsec\,pixel$^{-1}$ and 45\,s \& 720\,s Helioseismic and Magnetic Imager \citep[HMI,][]{2012SoPh..275..207S} high-resolution observations. The Solar Orbiter Extreme-Ultraviolet Imager \citep[EUI,][]{2020AA...642A...8R} high-resolution 174\,{\AA } EUV observations HRI\_EUV\_174 (0.5\,arcsec\,pixel$^{-1}$  \& 5\,s) are not available for the examined event. The Full Sun Imager (FSI) instrument aboard Solar Orbiter provided the EUV images at 174\,{\AA} and 304\,{\AA} with a time resolution of 10\,minutes and spatial resolution of 4.5\,arcsec\,pixel$^{-1}$. We examined radio spectrograms for event-associated nonthermal type III radio bursts from the following instruments: Solar Orbiter RPW \citep{2020AA...642A..12M}, Parker Solar Probe (PSP) FIELDS \citep{2016SSRv..204...49B}, Wind WAVES \citep{1995SSRv...71..231B} and STEREO-A WAVES \citep{2008SSRv..136..487B}.


\begin{figure*}
\center
\includegraphics[width=0.92\textwidth]{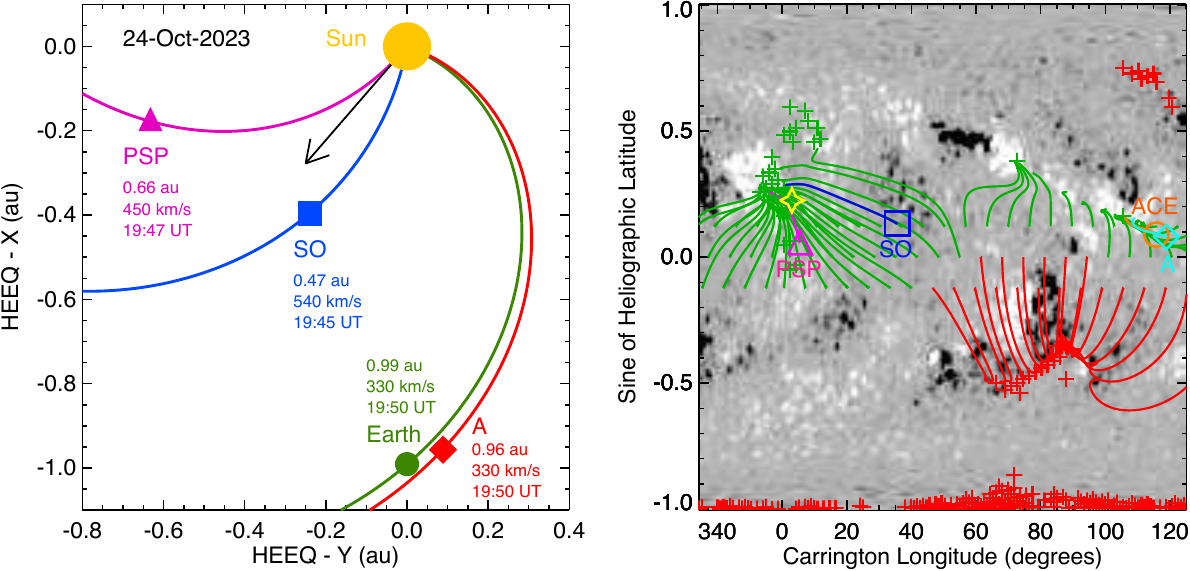}
\caption{Left: Location of PSP, Solar Orbiter (SO), Earth, and STEREO-A (A) in Heliocentric Earth Equatorial (HEEQ) coordinates at the time of the event-associated type III radio burst onset observed on each spacecraft. The arrow indicates the SEP source flare. The Parker IMF lines were derived from measured 1-hour averaged solar wind speeds at type III radio burst onset. For PSP, where solar wind speed is not available, we averaged values measured $\sim$1.5\,days before and $\sim$2 days after the type III burst. Heliocentric distance in au and solar wind speed in km\,s$^{-1}$ for respective spacecraft/locations are indicated. Right: Photospheric magnetic field map (grayscale), scaled to $\pm$30\,G for contrast enhancement, and PFSS model of an open coronal field (red is the negative and green is the positive magnetic polarity). Shown are field lines that intersect the source surface at latitudes 0$^{\circ}$ and $\pm$7$^{\circ}$. Upward triangle, square, circle, and diamond mark PSP, Solar Orbiter, near-Earth Advanced Composition Explorer (ACE), and STEREO-A magnetic footpoints at the source surface (set at 2.5\,R$_{\sun}$ from Sun center), respectively. The star marks the SEP source flare.
\label{fig:loc}}
\end{figure*}

\subsection{Magnetic connection} \label{subsec:mag}

Figure~\ref{fig:loc}(left) shows the X--Y plane projection of selected spacecraft in Heliocentric Earth Equatorial (HEEQ) coordinates at the 2023 October 24 event-associated type III radio burst onset. Solar Orbiter observed a type III burst at 19:45\,UT, PSP at 19:47\,UT, and near-Earth Wind and STEREO-A at 19:50\,UT. The type III burst was first visible at approximately 3 MHz at all four spacecraft. Figure~\ref{fig:loc}(right) displays a photospheric map along with a potential field source surface (PFSS) coronal field model on 2023 October 24 at 18:03:28\,UT, determined using the SolarSoft \citep{1998SoPh..182..497F} {\tt pfss} package\footnote{\url{https://www.lmsal.com/~derosa/pfsspack/}}. In the package, the model is sampled every 6\,hours. The photospheric map is generated from HMI magnetograms assimilated into the flux-dispersal model to provide the magnetic field on the full solar sphere \citep{2003SoPh..212..165S}. Figure~\ref{fig:loc}(right) shows that PSP and Solar Orbiter are connected to the coronal hole, via open (positive polarity) coronal field lines, where a nearby $^3$He-rich SEP source flare is located. We note that the IMF azimuthal angles on Solar Orbiter were between 270$^{\circ}$  and 360$^{\circ}$  (not shown) indicating away or positive polarity, i.e., consistent with PFSS model. Section~\ref{subsec:sol} discusses the identification of the solar source. Unfortunately, the PSP energetic particle instrument IS$\sun$IS \citep{2017JGRA..122.1513H} experienced a data gap on October 25 from 04 to 23\,UT, when Solar Orbiter observed a substantial part of the event.

\subsection{Energetic particles} \label{subsec:ene}

\begin{figure}
\center
\includegraphics[width=0.79\textwidth]{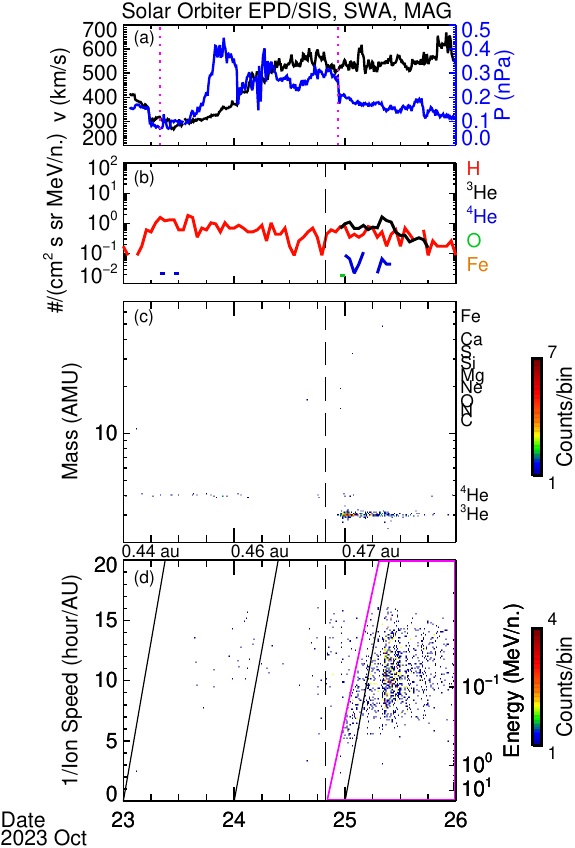}
\caption{Solar Orbiter measurements. (a) 10-min solar wind speed (black) and total pressure (blue). A pair of vertical magenta dotted lines approximately marks CIR. (b) 1-hour ion intensities at 0.23--0.32\,MeV\,nucleon$^{-1}$. (c) Mass versus time at 0.4--10\,MeV\,nucleon$^{-1}$. (d) 1/(ion speed) versus arrival times of 10--70\,AMU ions. Slanted lines mark arrival times for particles traveling along the nominal Parker field line without scattering. The trapezoid, highlighted in magenta, marks the period where energy spectra and elemental ratios were calculated. Heliocentric distances in au at beginning of each day are displayed. The measurements in panel (b) are from both SIS telescopes and panels (c--d) from SIS-a. The vertical black dashed line indicates the onset of the type III radio burst at Solar Orbiter (19:45\,UT).
\label{fig:par}}
\end{figure}

Figure~\ref{fig:par} plots elemental composition measurements and context solar wind data from Solar Orbiter for the 2023 October 24 $^3$He-rich event. Enhanced total pressure during the gradual solar wind speed increase (from $\sim$300 to $\sim$600\,km\,s$^{-1}$ in 1.5 days), a typical characteristic of Corotating Interaction Regions \citep[CIRs;][]{1995ISAA....3.....B}, is seen in Figure~\ref{fig:par}a (marked by dotted magenta lines). Furthermore, the gradual change in flow direction from west to east, another CIR characteristic, started at 19:30\,UT on October 23, around the beginning of the solar wind speed increase (not shown). The total pressure $P$ is given by the sum of the plasma and magnetic field pressure, i.e., $P = 2n_{\rm p}kT_{\rm p} + B^2/2\mu_0$, where $n_{\rm p}$ and $T_{\rm p}$ are the proton density and temperature, respectively, and $B$ is the magnetic field magnitude. In Fig.~\ref{fig:par}b, the H intensity does not show any rise in coincidence with the $^3$He, although the $^4$He does. So, the H is not due to the $^3$He injection. The panel shows that the H intensity increase started at $\sim$03 UT on October 23 in association with CIR and may dominate the $^3$He-rich SEP event during October 24--25. CIR ion intensity enhancements typically extend beyond CIR trailing edges and may last several days \citep[e.g.,][]{2009AnGeo..27.3677B,2012ApJ...749...73E}. Remarkably, the mass versus time spectrogram at energy 0.4--10\,MeV\,nucleon$^{-1}$ in Fig.~\ref{fig:par}c shows almost exclusively $^3$He, with only a few other ion species measured above energy $\sim$0.4\,MeV\,nucleon$^{-1}$. A velocity dispersion pattern on October 24 in Fig.~\ref{fig:par}d suggests the ion solar release time coincided with a type III burst at 19:45\,UT. The onset of the event at PSP was not measured. Either the onset moved to the data gap because of the particle’s later arrival at PSP (e.g., one hour for 0.4\,MeV\,nucleon$^{-1}$), or the event was too small to be measured at further distances from the Sun.

\begin{figure*}
\center
\includegraphics[width=.92\textwidth]{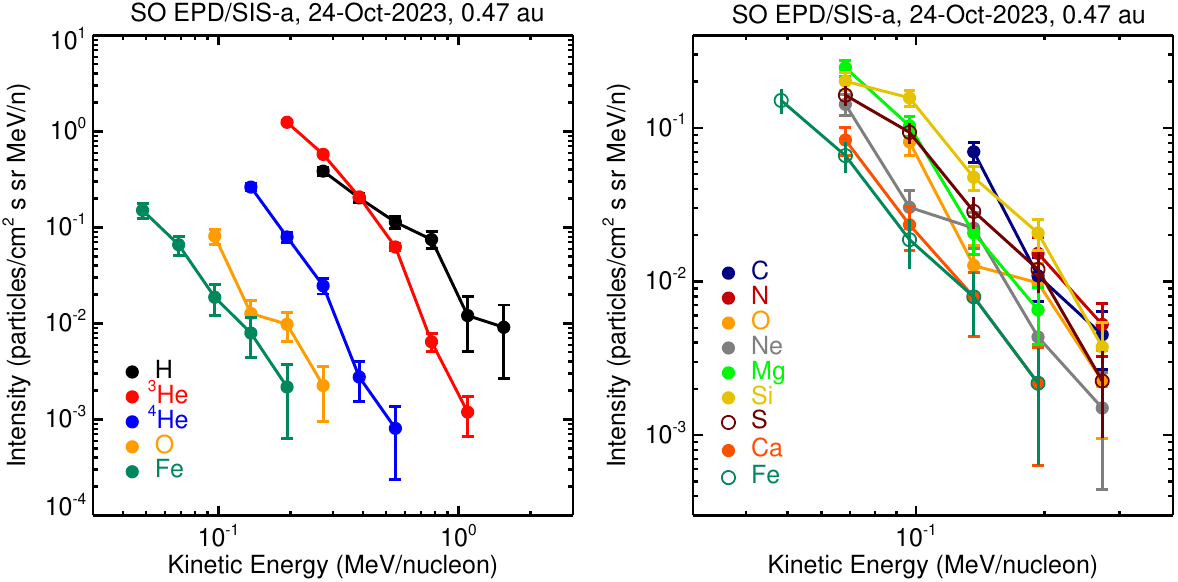}
\caption{Intensity spectra for selected ion species measured by SIS-a at the 2023 October 24 event. 
\label{fig:spe}}
\end{figure*}

Figure~\ref{fig:spe} shows intensity energy spectra for various ion species for the 2023 October 24 event. We focus on heavy ions, as H is likely dominated by the CIR event. The heavy-ion spectra are similar to each other with a power law-like shape. The $^3$He spectrum faintly bends toward low energies below $\sim$0.5\,MeV\,nucleon$^{-1}$, bringing the $^3$He/$^4$He ratio to increase with energy. Figure~\ref{fig:spe}(left) shows that $^3$He (except H, which is most likely all CIR) significantly dominates all species, and O prevails over Fe. The $^3$He looks similar to Fe and shifted to higher energy by a factor of $\sim$7, larger than the average shift factor of 3$\pm$1.3 in \citet{2024ApJ...974...54M}. An unusual feature seen in Figure~\ref{fig:spe}(right) is the dominance of lighter species (C, N, Ne, Mg, Si, S) over Fe. Interestingly, while $^3$He extends to $\sim$1\,MeV\,nucleon$^{-1}$, heavy ions (C--Fe) show spectral points below $\sim$0.2--0.3\,MeV\,nucleon$^{-1}$. Because C and N are not well resolved at low energies, their low-energy spectral points are not displayed. 

\begin{figure}
\center
\includegraphics[width=0.48\textwidth]{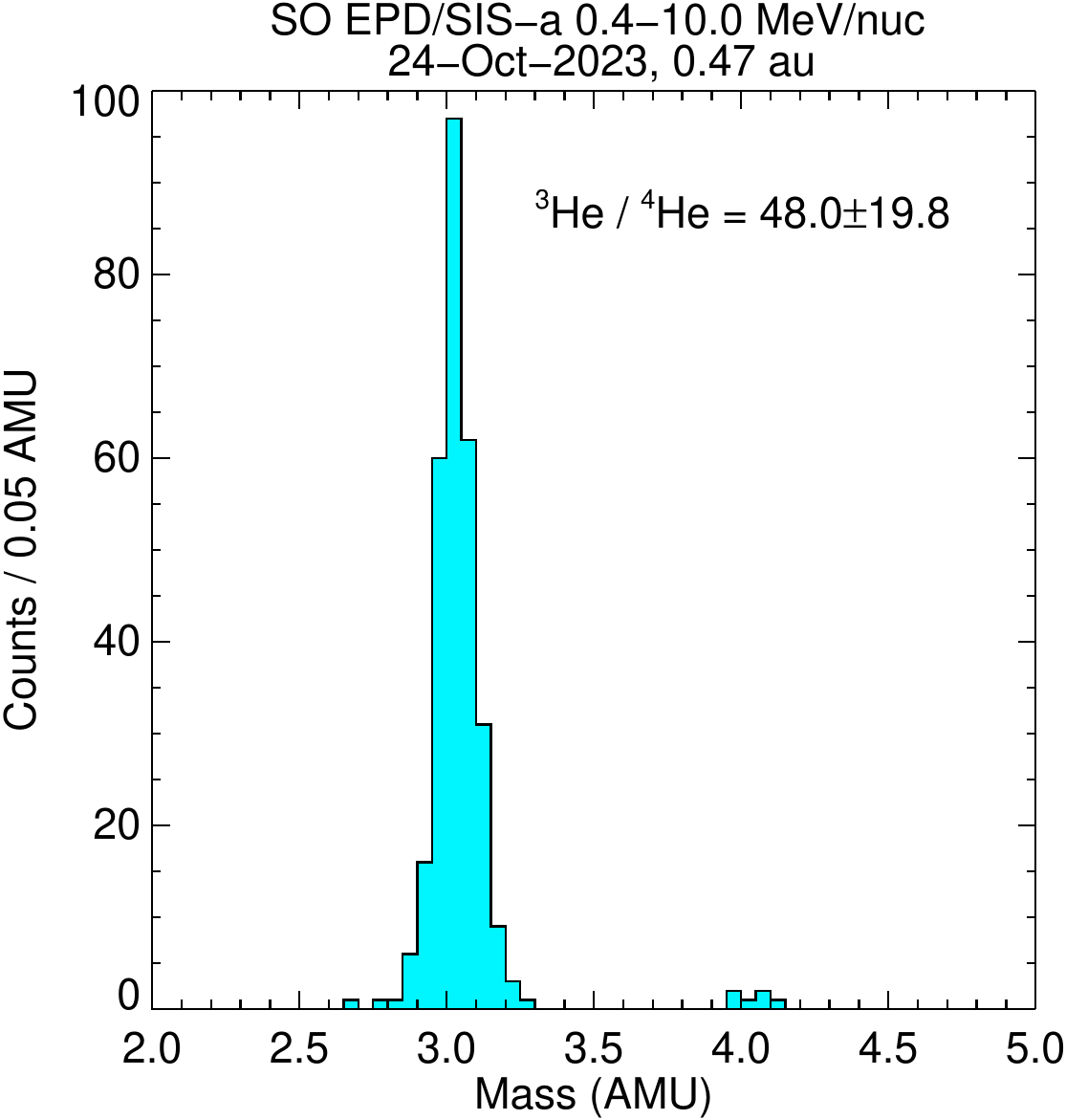}
\caption{SIS-a He mass histogram for the 2023 October 24 event in the energy range of 0.4--10.0\,MeV\,nucleon$^{-1}$.
\label{fig:his}}
\end{figure}

Figure~\ref{fig:his} shows the SIS-a He histogram for the examined event at 0.4--10.0\,MeV\,nucleon$^{-1}$ in the trapezoid range shown in magenta color in Fig.~\ref{fig:par}d. The $^3$He/$^4$He is 48.0$\pm$19.8 at 0.4--10.0\,MeV\,nucleon$^{-1}$ with a relatively high error (41\%) owing to a small number (six) of $^4$He counts. Note that in the same energy range and integration period, only four ions at 10--70 AMU were detected (cf. Fig.~\ref{fig:par}c). The ratio remains unchanged after decreasing the upper energy value to 3.0\,MeV\,nucleon$^{-1}$. The $^3$He/$^4$He at 0.1600--0.2263, 0.2263--0.3200, and 0.3200--0.4525\,MeV\,nucleon$^{-1}$ is 15.67$\pm$1.95, 23.39$\pm$4.29, and 75.20$\pm$33.85. The latter extremely high ratio is $\sim$1.8$\times$10$^{5}$ times the slow solar wind value of (4.08$\pm$0.25)$\times$10$^{-4}$ \citep{1995SSRv...72...61B,1998SSRv...84..275G}.

\begin{figure*}
\center
\includegraphics[width=.92\textwidth]{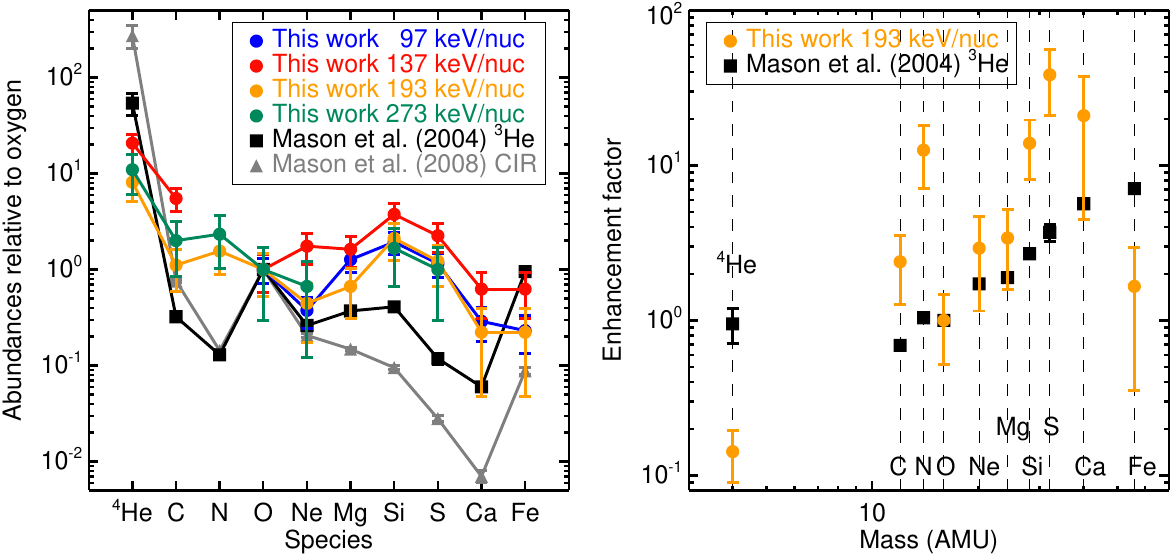}
\caption{Left: SIS-a 0.097, 0.137, 0.193, and 0.273\,MeV\,nucleon$^{-1}$ abundances relative to oxygen for the 2023 October 24 event (circles). Overplotted are 0.386\,MeV\,nucleon$^{-1}$ ion abundances from the $^3$He-rich SEP event survey by \citet{2004ApJ...606..555M} and the CIR event survey by \citet{2008ApJ...678.1458M}. Right: Abundance enhancement factor relative to oxygen for the 2023 October 24 event (at 0.193\,MeV\,nucleon$^{-1}$) relative to coronal abundances from \citet{1995AdSpR..15g..41R}. Black squares are values from \citet{2004ApJ...606..555M} $^3$He-rich SEP survey at 0.386\,MeV\,nucleon$^{-1}$.
\label{fig:abu}}
\end{figure*}

\begin{deluxetable*}{ccccccccc}
\tabletypesize{\scriptsize}
\tablecaption{Abundances and enhancement factors.\label{tab:t1}}
\tablehead{
\colhead{Element} & \multicolumn{6}{c}{Abundances} & \multicolumn{2}{c}{Enhancement factors} \\
\cline{2-7}
\colhead{} & \multicolumn{4}{c}{2023 Oct 24} & \colhead{2014 May 16$^{*}$} & \colhead{\citet{2004ApJ...606..555M} } & \colhead{2023 Oct 24} & \colhead{\citet{2004ApJ...606..555M} } \\
\cline{2-5}
\colhead{} & \colhead{0.097$^{\dagger}$} & \colhead{0.137$^{\dagger}$} & \colhead{0.193$^{\dagger}$} & \colhead{0.273$^{\dagger}$}  &  \colhead{0.4--1.0$^{\dagger}$} & \colhead{0.386$^{\dagger}$} & \colhead{0.193$^{\dagger}$}  & \colhead{0.386$^{\dagger}$} 
} 
\startdata 
$^4$He	& \nodata	& 20.7$\pm$5.0	 & 8.12$\pm$2.95 & 11.0$\pm$4.9 &	79.53$\pm$9.70&54$\pm$14 & 0.14$\pm$0.05 & 0.95$\pm$0.24\\
C&	 \nodata&	5.50$\pm$1.51	&1.11$\pm$0.52&	2.00$\pm$1.15 & 1.13$\pm$0.17&0.322$\pm$0.003	&2.40$\pm$1.16&	0.69$\pm$0.01\\
N	& \nodata&	 \nodata&	1.56$\pm$0.68&	2.33$\pm$1.30 &0.97$\pm$0.15	&0.129$\pm$0.002	&12.5$\pm$5.5	&1.04$\pm$0.03\\
O&	$\equiv$1.0$\pm$0.29	&$\equiv$1.0$\pm$0.42&	$\equiv$1.0$\pm$0.48&	$\equiv$1.0$\pm$0.71& $\equiv$1.0$\pm$0.15&	$\equiv$1.0$\pm$0.006	&$\equiv$1.0$\pm$0.48&	$\equiv$1.0$\pm$0.01\\
Ne&	0.38$\pm$0.14	&1.75$\pm$0.62	&0.44$\pm$0.27	&0.67$\pm$0.54 &0.58$\pm$0.10	&0.261$\pm$0.003	&2.92$\pm$1.77	&1.72$\pm$0.05\\
Mg&	1.27$\pm$0.35	&1.63$\pm$0.58&	0.67$\pm$0.36&	 \nodata	&0.61$\pm$0.11&0.370$\pm$0.003	&3.40$\pm$1.81&	1.89$\pm$0.04\\
Si&	1.93$\pm$0.50	&3.75$\pm$1.10	&2.11$\pm$0.87&	1.67$\pm$1.00&3.68$\pm$0.45	&0.409$\pm$0.004&	13.9$\pm$5.7&	2.69$\pm$0.07\\
S&	1.15$\pm$0.32	&2.25$\pm$0.74&	1.22$\pm$0.56	&1.00$\pm$0.71&3.83$\pm$0.46&	0.118$\pm$0.015	&38.4$\pm$17.6	&3.70$\pm$0.47\\
Ca&	0.29$\pm$0.11	&0.63$\pm$0.31&	0.22$\pm$0.17	& \nodata &0.45$\pm$0.18&	0.060$\pm$0.003&	21.0$\pm$16.5	&5.66$\pm$0.38\\
Fe&	0.23$\pm$0.10	&0.63$\pm$0.31&	0.22$\pm$0.17&	 \nodata	&3.00$\pm$0.37&0.950$\pm$0.005&	1.66$\pm$1.30	&7.09$\pm$0.21\\
\enddata
\tablecomments{$^{*}$Event in \citet{2016ApJ...823..138M}, $^{\dagger}$MeV\,nucleon$^{-1}$}
\end{deluxetable*}

Figure~\ref{fig:abu}(left) displays relative abundances for the 2023 October 24 event and $^3$He-rich SEP \citep{2004ApJ...606..555M} and CIR \citep{2008ApJ...678.1458M} reference elemental composition. In the examined event, the abundances at 0.137\,MeV\,nucleon$^{-1}$ are systematically higher by a factor of $\sim$2--3 compared to abundances at 0.097, 0.193, and 0.273\,MeV\,nucleon$^{-1}$ (when available) for all species except for C and Ne, whose 0.137\,MeV\,nucleon$^{-1}$ abundances are a factor of $\sim$4--5 higher. The abundances at 0.097, 0.193, and 0.273\,MeV\,nucleon$^{-1}$ for $^4$He, Si, S, Ca, and Fe differ within a factor of one between these three energies and for C, N, Ne, and Mg within a factor of two. The abundances of $^4$He and Fe are lower than the reference values reported by \citet{2004ApJ...606..555M}. The abundances of all other species show higher values than the $^3$He-rich SEP reference composition, although 0.193\,MeV\,nucleon$^{-1}$ Ca, Mg, and Ne, and 0.273\,MeV\,nucleon$^{-1}$ Ne are within statistical uncertainties. Although the error is large, the elevated C/O ratio is puzzling. It is more characteristic of CIR or solar wind abundances \citep{1991ApJ...382L..43R,2008ApJ...678.1458M}. The two species, $^4$He and C, have higher abundances in CIR events than in $^3$He-rich SEP events. Since $^4$He abundance in the event is smaller than typical $^3$He-rich SEP abundances, it is unlikely that the CIR contributes just to the C abundance enhancement. 

\begin{figure*}
\begin{interactive}{animation}{movie.mp4}
\includegraphics[width=.9\textwidth]{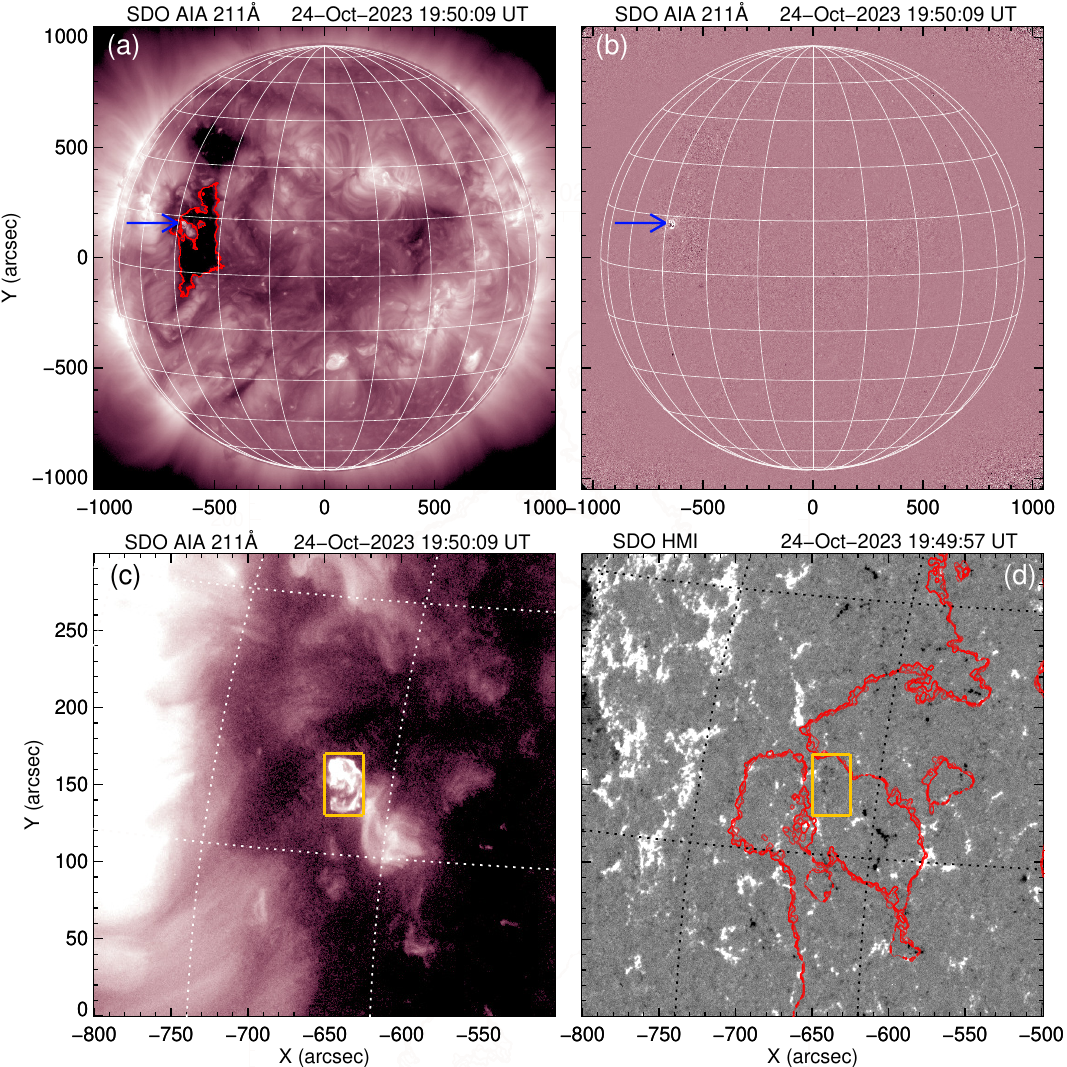}
\end{interactive}
\caption{(a) Direct and (b) 1-minute running difference SDO AIA EUV 211\,{\AA} images of the solar disk. Arrows point to the solar source of the $^3$He-rich SEPs in the 2023 October 24 event. The red contour marks a coronal hole near the solar source. (c) Zoom-in view of the area around the solar source in the direct AIA EUV 211\,{\AA}  image and (d) the SDO 45\,s HMI line-of-sight magnetogram scaled to $\pm$100\,G. The red contour of the coronal hole is overplotted. The yellow rectangles enclose the solar source. Images are at the time of the event-associated type III radio burst onset. Stonyhurst grid has 15$^{\circ}$ (10$^{\circ}$) spacing in a--b (c--d) panels. (An animation at 19:45:09--19:58:09\,UT corresponding to panel (b) and 19:44:57--19:59:45\,UT corresponding to panel (c) is available.)  
\label{fig:sol}}
\end{figure*}

Figure~\ref{fig:abu}(right) plots the abundance enhancement factor (X/O)/(X/O)$_{\text{corona}}$ of element X for the investigated $^3$He-rich SEP event at 0.193\,MeV\,nucleon$^{-1}$ (where abundances for all species are available) versus mass. The coronal abundances are represented by 5--12\,MeV\,nucleon$^{-1}$ gradual SEP abundances taken from \citet{1995AdSpR..15g..41R}. All abundances are normalized to oxygen. Shown are the average values from \citet{2004ApJ...606..555M} survey of 20 $^3$He-rich SEP events, where enhancements are also against gradual SEP abundances from \citet{1995AdSpR..15g..41R}. The enhancement pattern in the 2023 October 24 event is irregular. It does not show a smooth rise with mass. In addition, the enhancement factor through Mg--Si--S increases steeply with mass, in a factor of 10$+$, compared to a factor of 2 in a typical $^3$He-rich SEP event. A notable feature is the large enhancement of N, Si, and S and a decline in Fe and $^4$He. Table~\ref{tab:t1} gives abundances and enhancement factors for the examined event and \citet{2004ApJ...606..555M} survey, shown in Figure~\ref{fig:abu}. The abundances of the best-observed event (2014 May 16) from \citet{2016ApJ...823..138M} survey of $^3$He-rich SEP events with atypical enhancement patterns are listed.

\subsection{Solar source} \label{subsec:sol}

\begin{figure*}
\center
\includegraphics[width=0.93\textwidth]{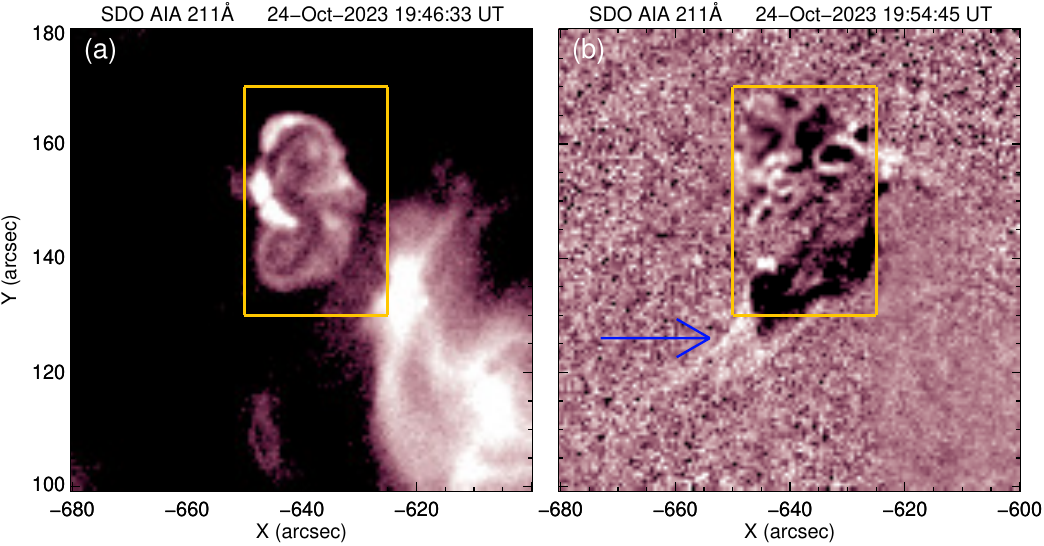}
\caption{(a) Direct AIA EUV 211\,{\AA} image of a mini filament (dark letter S-like shape, whose upper segment is bounded from outside by a brightening) in the solar source. (b) 1-minute running difference AIA EUV 211\,{\AA} image of a jet (marked by an arrow) ejected from the solar source. The yellow rectangles enclose the solar source.
\label{fig:sol1}}
\end{figure*}

Figure~\ref{fig:sol} displays the solar source of the investigated event in the AIA 211\,{\AA} image and 45\,s HMI line-of-sight (LOS) magnetogram at the time of the event-associated type III radio burst (as observed at 1\,au where the images were obtained). The solar source (a bright point), marked by arrows in Figs.~\ref{fig:sol}a--b, is located near the edge of the low-latitude coronal hole (dark area in the AIA 211\,{\AA} direct image). The coronal hole boundary, highlighted in red contours, was obtained from the SolarSoft Collection of Analysis Tools for Coronal Holes \citep[CATCH,][]{2019SoPh..294..144H}. Stonyhurst’s heliographic longitude and latitude of the solar source are $-$43$^{\circ}$ (E43) and 13$^{\circ}$ (N13), respectively. Note that the Carrington longitude of the source is 3$^{\circ}$ (see Fig.~\ref{fig:loc}(right)). The solar source was identified using a standard approach based on the temporal coincidence between EUV brightening observed on the solar disk and the event-associated type III radio burst. Fig.~\ref{fig:sol}a--b shows that the only brightening was one near the coronal hole (see also the animation corresponding to Fig.~\ref{fig:sol}b). The animation, related to Fig.~\ref{fig:sol}c, reveals an eruption of the dark (cool) mini filament of a letter S-like shape at around 19:48\,UT. Simultaneously, the brightness at the source notably increases. This is followed by a jet ejection observed at around 19:53\,UT in the southeast direction. Except for 211\,{\AA}, the mini filament was observed in 193, 171, 131, and 304\,{\AA} wavelengths. The yellow rectangle enclosing the solar source (and base of the jet) has a diagonal of 34\,Mm, the size of a supergranulation scale \citep[$\sim$30 Mm;][]{2010LRSP....7....2R}. The bright area, a plage, eastward from $-$50$^{\circ}$, corresponds to relatively high concentrations of strong positive (white) polarity magnetic fields in the LOS magnetogram (see Fig.~\ref{fig:sol}c--d). The remaining area on the west contains weak (grey) magnetic fields of the coronal hole with several magnetic elements, mostly of positive polarity. The unsigned (total) mean magnetic field strength in the rectangle surrounding the source is $\sim$8\,G, significantly lower than fields in active regions (ARs). It was obtained by averaging absolute values of the magnetic field strength of each pixel (there were 4000 pixels in the rectangle). We use the 720\,s magnetogram at 19:58\,UT (the closest one to the type III radio burst) due to lower photon noise than 45\,s magnetograms. After correction for angular distance from the center of the solar disk, the value is $\sim$11\,G. Figure~\ref{fig:sol1}a shows a mini filament in the source of the October 24 event about 1 minute before its eruption, and Fig.~\ref{fig:sol1}b shows a weak straight jet that has a size similar to its base. Previous studies have reported the eruption of the mini filament as a trigger of coronal jets \citep[e.g.,][]{2016ApJ...821..100S,2016ApJ...832L...7P}.

No GOES X-ray flare was observed in association with this event. We note that the C1.3 GOES X-ray flare at 19:53\,UT and a maximum at 20:18\,UT occurred on the east limb (S13E88). The solar source of the investigated event was coincidentally captured by the Interface Region Imaging Spectrograph \citep[IRIS,][]{2014SoPh..289.2733D} observatory in its limited field-of-view (165$\times$175\,arcsec$^{2}$), in Si IV 1400\,{\AA} and Mg II k 2796\,{\AA} slit-jaw UV images. However, spectra cannot be obtained around the time of the type III radio bursts since the slit passed through the source at 20:13\,UT, i.e., when the eruption had already ceased. 

\subsection{Solar source temperature} \label{subsec:tem}

\begin{figure*}
\center
\includegraphics[width=.94\textwidth]{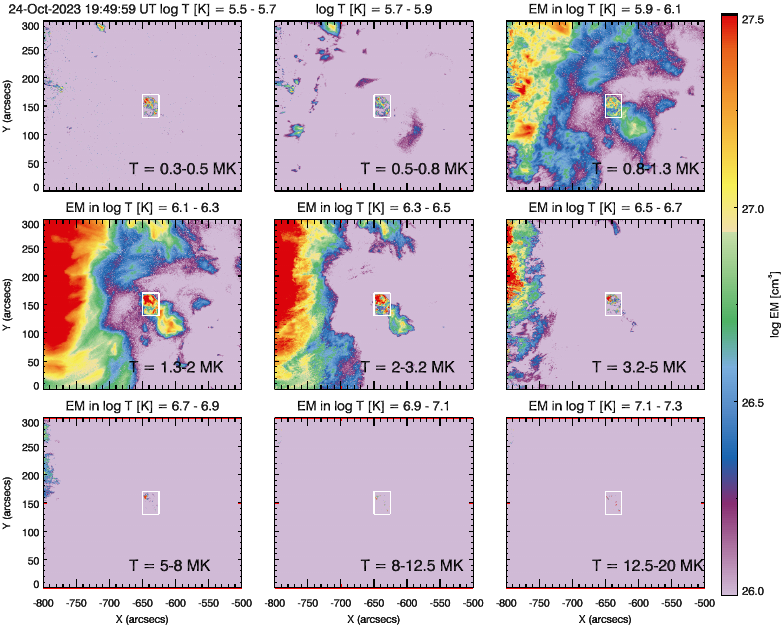}
\caption{Emission Measure (EM) maps for the 2023 October 24 event at the start time of the type III radio burst. The white rectangle surrounds the source region, where the temperature was determined. Color-coding indicates the total EM contained within a log~$T$ range, shown at the top of each panel. 
\label{fig:tem}}
\end{figure*}

Differential Emission Measure (DEM) was previously used to determine the temperature at the solar source in 24 $^3$He-rich SEPs events \citep{2021ApJ...908..243B}. The authors calculated the EM-weighted temperature at the time of the event-associated type III radio burst. This method provided a reasonable agreement with temperature obtained from in situ Fe charge states at $\sim$0.1\,MeV\,nucleon$^{-1}$ \citep{2008ApJ...687..623D} and the elemental abundance measurements \citep[e.g.,][]{2014SoPh..289.3817R}. Here we use EM to derive the temperature and its evolution at the solar source of the 2023 October 24 event. DEM is a physical quantity related to the electron number density and temperature gradient of a plasma. It is determined from observations of emission lines on the assumption that plasma is optically thin and in ionization equilibrium. The DEM is reconstructed from multi-wavelength observations and the instrumental temperature response by the inversion method. We use DEM inversion\footnote{\url{https://www.lmsal.com/~cheung/AIA/tutorial_dem/}} developed by \citet{2015ApJ...807..143C} for EUV AIA imaging observations. We coalign AIA images in six channels (94, 131, 171, 193, 211, and 335\,{\AA}), which are sensitive to a range of temperatures \citep{2010AA...521A..21O,2011AA...535A..46D,2013AA...558A..73D}, and use them as input in the EM analysis. We computed the AIA temperature response function using the technique given by \citet{2011AA...535A..46D}. We created an isothermal spectrum using the latest CHIANTI version \citep{2019ApJS..241...22D}, the electron number density of 1$\times$10$^{10}$\,cm$^{-3}$ and coronal abundances (sun\_coronal\_2021\_chianti.abund - available in the CHIANTI database). This spectrum was then folded with the effective area of each AIA filter (obtained using the SolarSoft routine aia\_get\_response.pro), and temperature responses were obtained. The EM (the integral of DEM) is computed for each pixel position in the solar source and its surrounding area, and the temperature interval log~$T$ (K)$=$ 5.4--7.5 (0.3--30\,MK). Finally, the EM is spatially averaged over the rectangular region around the source (shown as a white box in Fig.~\ref{fig:tem}). The region was selected to contain EUV brightening in running differences images at type III radio burst onset \citep{2021ApJ...908..243B}. The DEM inversion code gives temperature within an error of 20\%. The errors arise from instrumental effects and photon counting statistics.

Figure~\ref{fig:tem} shows EM maps at various temperature intervals at the solar source and its surroundings for the 2023 October 24 $^3$He-rich SEP event at the time of the event-associated type III radio burst. The solar source is bounded by a white rectangle, the same as in Fig.~\ref{fig:sol}c,d. We obtained such maps between 19:39\,UT and 20:05\,UT with a temporal resolution of 12\,s. 

\begin{figure*}
\center
\includegraphics[width=.97\textwidth]{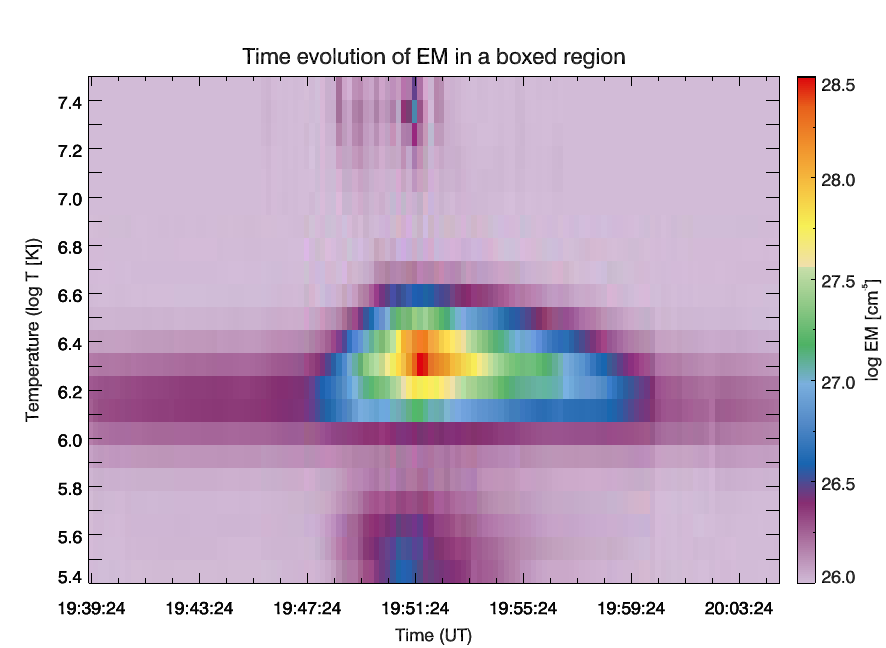}
\caption{Temperature versus time in the rectangle around the SEP source. The spatially averaged emission measure is color-coded. 
\label{fig:evl}}
\end{figure*}

Figure~\ref{fig:evl} shows color-coded EM in a temperature versus time diagram on 2023 October 24 from 19:39 to 20:05\,UT. The EM is spatially averaged in the area bounded by the rectangle in Fig.~\ref{fig:tem}. Between 19:39 and 19:47\,UT, 1.3\,MK (log~$T$ 6.1\,K) and 1.6\,MK (log~$T$ 6.2\,K) dominate the region as they have the highest and the same EMs. At 19:48\,UT (mini filament eruption), the EM starts increasing, but 1.3 and 1.6\,MK still dominate. At 19:49\,UT, 1.6 and 2.0\,MK (log~$T$ 6.3\,K) dominate. Between 19:50\,UT (type III burst onset) and 19:56\,UT, 2.0\,MK dominates the source region. Later, lower temperatures again prevail. At 19:50\,UT, the region also clearly shows temperatures around 0.3\,MK (log~$T$ 5.4 and 5.5\,K), though with a factor of three lower EMs than 2\,MK (see Fig.~\ref{fig:ftt}c).

\section{Discussion} \label{subsec:dis}

\subsection{Solar sources of previously reported similar events } \label{subsec:prev}

\begin{deluxetable*}{cccccccccc}
\tablecaption{Previously reported $^3$He-rich SEP events with abundance enhancement pattern not ordered by mass.\label{tab:t2}}
\tablehead{
\colhead{\#} & \colhead{SEP start date} &  \colhead{CH} &  \multicolumn{3}{c}{EUV flare} & \colhead{Mini filament} & \colhead{Type III} & \colhead{$^3$He/$^4$He$^{\dagger}$} &\colhead{log $T$$^{\rm f}$ (K)} \\
\cline{4-6}
\colhead{} & \colhead{} & \colhead{} & \colhead{type} & \colhead{location} & \colhead{time} &\colhead{}  &\colhead{}  &\colhead{} &\colhead{} 
} 
\colnumbers
\startdata 
1 &	1999-Mar-21$^{\rm a}$ & Y$^{\rm b}$  & Jet$^{\rm b}$  & N19W27$^{\rm b}$  &	04:36$^{\rm b}$  &	?$^{\rm c}$  &	?$^{\rm c}$  &	3.65$\pm$0.15& \nodata\\
2 &	2000-Jan-06$^{\rm a}$&	Y$^{\rm b}$&	Ejection$^{\rm b}$&	N15W52$^{\rm b}$&	02:36$^{\rm b}$&	?$^{\rm c}$&	02:03$^{\rm c}$&	23.42$\pm$4.69& \nodata\\
3& 	2000-Jan-09$^{\rm a,d}$& \nodata	& \nodata& \nodata& 	\nodata& 	\nodata	& 16:48$^{\rm c,*}$& 	0.29$\pm$0.02& \nodata\\
4&	2000-Feb-02$^{\rm d}$&	 \nodata&	\nodata	&\nodata&	\nodata&	\nodata&	09:45$^{\rm c}$&	\nodata& \nodata\\
5&	2001-Jul-16$^{\rm a,d}$	& \nodata	&\nodata&	\nodata	&\nodata&	\nodata&	11:53$^{\rm c}$&	0.60$\pm$0.09& \nodata\\
6&	2002-Apr-05$^{\rm a}$&	 \nodata	&\nodata&	\nodata&	\nodata&	\nodata&	20:42$^{\rm c,*}$ &	0.05$\pm$0.01& \nodata\\
7&	2002-Sep-25$^{\rm a}$	& \nodata	&\nodata &	\nodata &	\nodata &	\nodata	&18:42$^{\rm c,*}$ &	2.33$\pm$0.46& \nodata \\
8&	2002-Oct-20$^{\rm a}$	&Y$^{\rm b}$&	Jet$^{\rm b}$ &	S16W60$^{\rm b}$&	14:24$^{\rm b}$&	?$^{\rm c}$&	14:11$^{\rm c}$&	1.35$\pm$0.04& \nodata\\
9&	2002-Oct-21$^{\rm a}$	&Y$^{\rm b}$&	Jet$^{\rm b}$	& S16W68$^{\rm b}$&	04:24$^{\rm b}$	&?$^{\rm c}$&	04:17$^{\rm c}$&	1.28$\pm$0.05& \nodata\\
10&  2003-Apr-28$^{\rm a}$&	\nodata	&\nodata&	\nodata&	\nodata	&\nodata	&04:01$^{\rm c}$&	0.22$\pm$0.02& \nodata\\
11&	2011-Jul-9$^{\rm a}$	&Y$^{\rm e}$&	Jet$^{\rm e}$	& N14W44$^{\rm e}$&	16:25$^{\rm e}$&	Y$^{\rm c}$&	16:25$^{\rm e,*}$&	5.21$\pm$0.49& 6.36\\
12&	2012-Aug-02$^{\rm a}$	&N$^{\rm c}$&	Jet$^{\rm f}$	& N12W58$^{\rm f}$&	03:00$^{\rm f}$&	Y$^{\rm c}$&	03:00$^{\rm f}$&	0.45$\pm$0.02 & 6.32\\
13&	2012-Nov-20$^{\rm a}$	&N$^{\rm c}$&	Jet$^{\rm g,h}$&	S17W60$^{\rm g,h}$&	01:31$^{\rm g}$&	Y$^{\rm c}$&	01:30$^{\rm g,h}$&	7.66$\pm$0.87 & 6.31\\
14&	2014-Apr-17$^{\rm a}$	&N$^{\rm c}$&	Eruption$^{\rm g}$& 	S15W24$^{\rm g}$&	21:50$^{\rm g}$&	Y$^{\rm c}$&	21:58$^{\rm g}$&	0.61$\pm$0.04 & 6.54\\
15&	2014-Apr-24$^{\rm a}$&	Y$^{\rm c}$&	Eruption$^{\rm g}$	& S18W102$^{\rm g}$	&00:50$^{\rm g}$	&Y$^{\rm c}$&	00:40$^{\rm g}$&	1.13$\pm$0.06& \nodata\\
16&	2014-May-16$^{\rm a}$	&Y$^{\rm g}$&	Jet$^{\rm g}$&	 S12W44$^{\rm g}$&	04:02$^{\rm c}$&	Y$^{\rm a,i}$&	03:57$^{\rm g}$&	14.88$\pm$1.36 & 6.31\\
17&	2015-Feb-06$^{\rm a}$	&N$^{\rm c}$	&Jet$^{\rm f}$&	N14W19$^{\rm f}$&	23:31$^{\rm c}$&	Y$^{\rm c}$&	23:30$^{\rm f,*}$&	1.76$\pm$0.28 & 6.19\\
18&	2022-Mar-06$^{\rm j}$	&N$^{\rm c}$	&Jet$^{\rm j}$&	W26S16$^{\rm j}$&	23:55$^{\rm c}$&	Y$^{\rm j}$&	23:55$^{\rm j,*}$&	0.97$\pm$0.02 & \nodata\\
19&	2023-Apr-08$^{\rm k}$	&\nodata	 & \nodata & \nodata& \nodata&\nodata&	\nodata&	5.50$\pm$0.30 & \nodata\\
\enddata
\tablecomments{ 
$^{\rm a}$Event in \citet{2016ApJ...823..138M}, 
$^{\rm b}$\citet{2006ApJ...639..495W}, 
$^{\rm c}$This work, 
$^{\rm d}$Event in \citet{2002ApJ...565L..51M}, 
$^{\rm e}$\citet{2014ApJ...786...71B}, 
$^{\rm f}$\citet{2021ApJ...908..243B}, 
$^{\rm g}$\citet{2015ApJ...806..235N}, 
$^{\rm h}$\citet{2015AA...580A..16C}, 
$^{\rm i}$\citet{2016AN....337.1024I}, 
$^{\rm j}$Event in \citet{2023AA...673L...5B},
$^{\rm k}$Event in \citet{2023ApJ...957..112M}, 
$^{*}$Day before SEP start date, 
$^{\dagger}$\#1--3, \#5--17 at 0.5--2.0\,MeV\,nucleon$^{-1}$ and \#18, \#19 at  0.320--0.452\,MeV\,nucleon$^{-1}$
} 
\end{deluxetable*}

\citet{2002ApJ...565L..51M,2016ApJ...823..138M} reported 17 events measured by ACE in 1997--2015 showing no simple heavy-ion enhancement pattern with mass with greatly enhanced C/O, N/O, Si/O, and S/O over average values in $^3$He-rich SEP events. Two other events were reported from Solar Orbiter observations \citep{2023AA...673L...5B,2023ApJ...957..112M}. For most of these events, spectra were measured below 1\,MeV\,nucleon$^{-1}$. Table~\ref{tab:t2} summarizes the sources of these events. Column 1 lists the event number. Column 2 provides the $^3$He-rich SEP event start date. Column 3 indicates with Y (yes) or N (no) the presence of a coronal hole (CH). Columns 4, 5, and 6 give EUV flare type, location, and time, respectively. The EUV refers to 195\,{\AA} SOHO Extreme-ultraviolet Imaging Telescope \citep[EIT,][]{1995SoPh..162..291D} for \#1--2 and \#8--9, 193 Å SDO AIA for \#11--14 and \#16--18, and 195\,{\AA} STEREO-A Extreme Ultraviolet Imager \citep[EUVI,][]{2008SSRv..136...67H} for \#15. Column 7 indicates the observation of an erupting mini filament in EIT 195\,{\AA} (SOHO 304\,{\AA} images were not available) or AIA (EUVI) 304\,{\AA}. Column 8 provides the start time of type III radio bursts from Wind WAVES (\#2, \#4--17), STEREO-A WAVES (\#18), and the US National Oceanic and Atmospheric Administration (NOAA) Edited Events catalog (\#3). Column 9 lists a $^3$He/$^4$He ratio at 0.5--2.0\,MeV\,nucleon$^{-1}$ (\#1--3, \#5--17) and 0.32--0.45\,MeV\,nucleon$^{-1}$ (\#18, \#19). In the 2023 October 24 event, the $^3$He/$^4$He at 0.5--2.0\,MeV\,nucleon$^{-1}$ is 28.5$\pm$14.5. The last column provides the EM-weighted temperatures.

\begin{figure*}
\center
\includegraphics[width=0.94\textwidth]{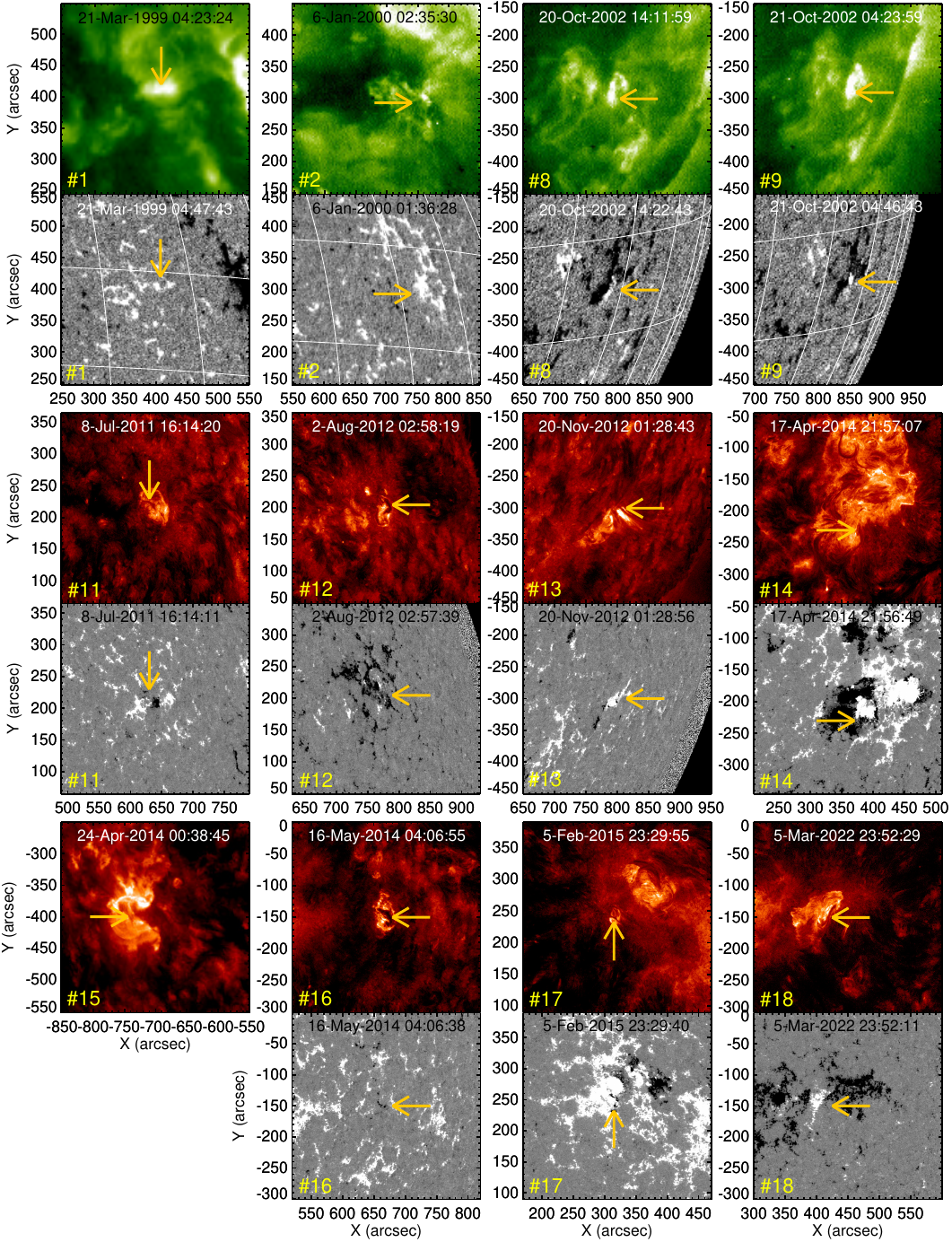}
\caption{Solar sources of events \#1--2 and \#8--9 in EIT 195\,{\AA} images and MDI magnetograms. For these events, arrows mark EUV flare location from Table~\ref{tab:t2} (column 5). Solar sources of events \#11--14 and \#16--17 in AIA 304\,{\AA} images and HMI magnetograms. Solar source of event \#15 in EUVI 304\,{\AA}. For events \#11--18, arrows mark location of erupting mini filaments.
\label{fig:old}}
\end{figure*}

Twelve events in Table~\ref{tab:t2} had sources previously identified. These solar sources are displayed in Fig.~\ref{fig:old}; \#1--2, \#8--9 in SOHO images, \#11--14, \#16--18 in SDO and \#15 in STEREO-A images. The solar sources of four events \#1--2 and \#8--9 were inspected by \citet{2006ApJ...639..495W} with SOHO EIT (2.6\,arcsec\,pixel$^{-1}$ \& 12 minutes) 195\,{\AA} images and nearby CHs with near-infrared He I 10830\,{\AA} data from National Solar Observatory Kitt Peak. \citet{2006ApJ...639..495W} explored $^3$He-rich SEP events from 1997--2003 whose sources the authors were able to identify. Table~\ref{tab:t2} includes ten events in this period, but only four of them have been reported by \citet{2006ApJ...639..495W}. Sources of events \#3--7 and \#10 were presumably too small to be resolved by EIT. The same applies to mini filament identification in events \#1--2 and \#8--9, though some hints of narrow dark features, resembling a mini filament, could be observed in sources for events \#2 and \#8. For event \#1, only half spatial resolution EIT images are available. The photospheric field underneath the solar sources in events \#1--2 and \#8--9 is shown in SOHO 2\,arcsec\,pixel$^{-1}$ \& 96 minute cadence Michelson Doppler Imager \citep[MDI,][]{1995SoPh..162..129S} magnetograms. The mini filaments in events \#16 \citep{2015ApJ...806..235N,2016AN....337.1024I,2016ApJ...823..138M} and \#18 \citep{2023AA...673L...5B} were identified in earlier studies. Event \#18 in Table~\ref{tab:t2} (event \#5 in \citet{2023AA...673L...5B}) was one of the six recurrent $^3$He-rich SEP events measured on March 3--6 by Solar Orbiter at $\sim$0.5\,au. Remarkably, the recurrent event with the mini filament eruption was rich in Si--S. Event \#19, observed by Solar Orbiter at 0.29\,au, had an uncertain solar source. The solar source in event \#14 is noticeably different -- it is larger and hotter. The arrow in the source for this event points to the faint dark feature (possible mini filament). This is also a site with fast changes in the brightening. The photospheric field underneath the solar sources for events \#11--14 and \#16--17 is displayed in HMI magnetograms. We note that event \#15 with a source behind the limb, observed by STEREO-A, has no magnetogram observations (therefore blank space in Fig.~\ref{fig:old}). A mini filament in this event was inspected with 304\,{\AA} EUVI images (1.6\,arcsec\,pixel$^{-1}$ \& 2.5 minutes).

The underlying photospheric field for events in Fig.~\ref{fig:old} is fairly variable. We observe relatively small magnetic elements, suggesting a weak surface field (class I), in events \#1, \#2, and \#16. The 2023 October event is in this category. Bipolar ARs (class II) are seen in events \#8, \#9, \#11, and \#12 \citep[see][for \#11]{2014ApJ...786...71B}. A plage region (class III) was previously reported in event \#13 \citep{2015AA...580A..16C} and \#18 \citep{2023AA...673L...5B}. Finally, large sunspots with magnetic moving features of opposite polarity (class IV) are seen in events \#14 and \#17. Remarkably, the events with the highest $^3$He/$^4$He ratios (\#2 and \#16) have a class I solar source, while the lowest $^3$He/$^3$He (except \#11) have class II and IV sources. Note that AR in event \#11 with a rather high $^3$He/$^4$He ratio was a newly emerging region within the weak field of CH \citep{2014ApJ...786...71B}. 

\subsection{Solar source of 2023 October 24 event} \label{subsec:oct}

\citet{2002ApJ...565L..51M,2016ApJ...823..138M} suggested that cool material must be present at the source, producing ionization states for selective C, N, Si, and S enhancement, and not getting temperatures high enough to bring Ca and Fe to the required ionization states. \citet{2016ApJ...823..138M} discussed the cool mini filament observed in event \#16. Remarkably, a mini filament was observed not only in the 2023 October event but virtually in all other sources in Table~\ref{tab:t2} when resolution allows identification. Furthermore, it has been suggested that such enhancement is produced by a range of temperatures occurring subsequently as the plasma is heated from initial low temperatures \citep[e.g.,][]{1980ApJ...239.1070M,2016ApJ...823..138M}. Specifically, a large enhancement of N and Si could be produced by resonant heating with the same waves of ionization states of these ions that occur at different temperatures \citep{2002ApJ...565L..51M}. Furthermore, \citet{1980ApJ...239.1070M} suggested that waves need to be generated repeatedly during the heating phase. Here, we show that various temperatures may coexist at a source of $^3$He-rich SEPs (Fig.~\ref{fig:evl}). It may suggest that common enhancement of lighter and heavier ion species would not necessarily need to be achieved subsequently but could happen in a single episode (or finite short time), e.g., during a stage when magnetic reconnection produces required waves. \citet{2023Univ....9..466R} argued that reconnection and particle acceleration occur rapidly, early, before much heating.

The 2023 October 24 event shows extremely large $^3$He enrichment. Event \#2 has the greatest $^3$He/$^4$He ratio in Table~\ref{tab:t2} \citep[33.4$\pm$5.2 at 0.385\,MeV\,nucleon$^{-1}$,][]{2000ApJ...545L.157M}, and event \#16 has the second highest $^3$He enrichment. A common feature of these three events is their surface magnetic field (class I). \citet{2006ApJ...636..462L} pointed out that situations with weak turbulence produce higher $^3$He enrichments than cases with strong turbulence \citep[see also Fig.~8 in][]{2024ApJ...974...54M}. The authors also concluded that in a strongly magnetized plasma, the $^3$He enhancement decreases. These model outcomes are qualitatively consistent with observations in these events on a small source size and weak, simple magnetic field configurations.

The fact that only $^3$He is observed above 0.5\,MeV\,nucleon$^{-1}$ in the 2023 October 24 event could be due to its unique $Q/M$ ratio (2/3 for $^3$He$^{+2}$) along with the small size of the jet where possibly only limited (wave) energy for ion acceleration is generated. $^3$He, with a unique $Q/M$, does not compete in absorbing wave energy with other ions and can spend a proportionally larger portion of available energy. Heavy ions in events \#3--5 in Table~\ref{tab:t2} were also observed at low ($<$0.5\,MeV\,nucleon$^{-1}$) energies; however, these events have unknown solar sources. They all have $^3$He/$^4$He and Fe/O typical of $^3$He-rich events \citep{2002ApJ...565L..51M,2016ApJ...823..138M}. Therefore, the examined event is special for its extreme $^3$He enrichment. Event \#16 (highly enriched in $^3$He) is the one in Table~\ref{tab:t2}, whose heavy-ion spectra extend above $\sim$1\,MeV\,nucleon$^{-1}$ \citep{2016ApJ...823..138M}. A weak photospheric magnetic field at the source of this event is a common feature with the 2023 October 24 event. However, event \#16 showed a large, twisted jet \citep{2015ApJ...806..235N,2016AN....337.1024I,2016ApJ...823..138M} with a projected size about eight times greater than the straight jet in the October 24 event, supporting an idea about a possible connection between jet size/complexity and energy attained by heavy ions. Note that the base of the jet in both these events has a similar scale size and is located at the same angular distance from the center of the solar disk, implying a similar projection effect. Further supporting observations can be found in \citet{2018ApJ...852...76B}. The authors reported three $^3$He-rich SEP events having energy spectra of heavy ions extending beyond $\sim$1\,MeV\,nucleon$^{-1}$ with sources at the coronal hole edge associated with twisted and extremely elongated jets (projected length spanning 40$^{\circ}$--60$^{\circ}$ of heliographic longitude). All jets were triggered by mini-filament eruptions. $^3$He/$^4$He and Fe/O ratios were $\sim$1--2. The heavy ion enhancement patterns were not investigated in these events.

\citet{1997ApJ...477..940R} discussed the energization details of $^3$He and the heavy ions. The authors argue that heavy ion energization can be significantly smaller because of their small acceleration rate. If the rate of ion acceleration is related to the jet, then a small and simple jet might not energize heavy ions efficiently. \citet{2004ApJ...613L..81L,2006ApJ...636..462L} point to the faster acceleration of $^3$He (or conversely, bigger losses for $^4$He) that let $^3$He attain more energy. Since the losses are sensitive to temperature, the $^4$He acceleration is suppressed as temperature decreases \citep{2006ApJ...636..462L}. It leads to a decrease in $^3$He/$^4$He with increasing temperature. Fitting the energy spectra of $^3$He and $^4$He, the authors reported the temperature for events \#1 and \#2 in Table~\ref{tab:t2}, which were 1.6 and 1.2\,MK, respectively. These temperatures are similar to the temperature of the examined event. \citet{2024ApJ...974...54M} suggested that the huge $^3$He/$^4$He ratios are due to a combination of higher energy achieved by the $^3$He along with a rollover in the spectrum at the high energies. 

The examined event showed a decline in $^4$He. \citet{2019SoPh..294...37R} reported rare $^4$He-poor events among $^3$He-rich events with abundances $^4$He/O$<$30. The author discussed that in a model of the FIP effect, $^4$He, with its unique high FIP value (24.6\,eV), is slowly ionized in transit to the corona \citep{2009ApJ...695..954L}, leading to its suppression by a factor of two. \citet{2019SoPh..294..141R} speculated that the observed sufficient $^4$He suppression may occur if the transition is fast, for example, in jets. \citet{2019SoPh..294...33H} reported four out of a hundred $^3$He-rich events with no $^4$He intensity increase above the background at $\sim$1\,MeV\,nucleon$^{-1}$ (though $^4$He could be present at lower energies).

\subsection{Temperature and heavy-ion elemental composition in the 2023 October 24 event } \label{subsec:comp}

\begin{figure}
\plotone{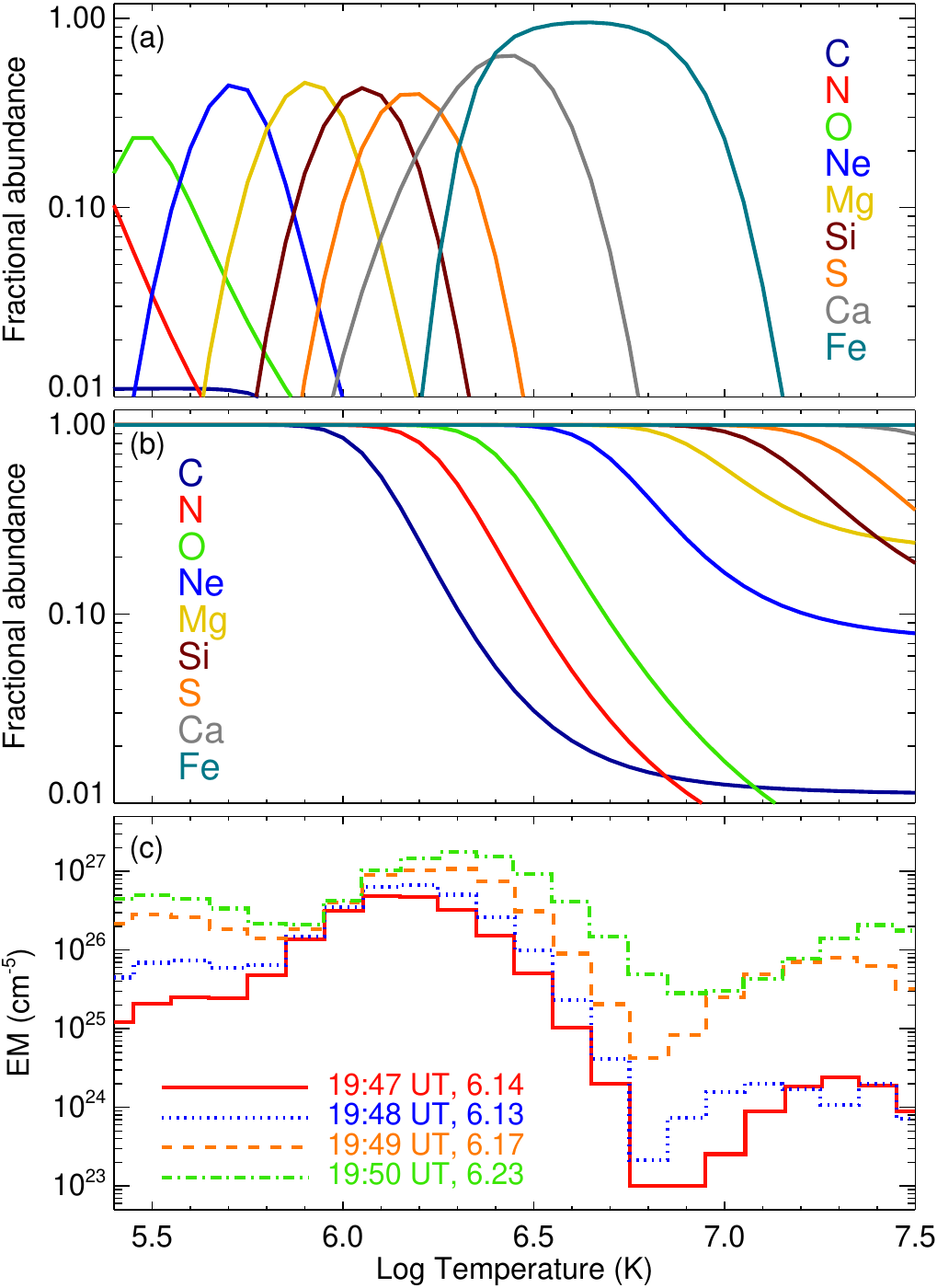}
\caption{(a) Fractional abundance of selected heavy-ion species whose gyrofrequency falls in the range of preferential heating given by formula (1) as a function of equilibrium temperature \citep[after][]{1980ApJ...239.1070M}. (b) Fractional abundance of selected heavy-ion species whose $Q/M$ falls below the damped wave region \citep[after][]{2023ApJ...957..112M} as a function of equilibrium temperature. (c) The EM profiles in the solar source (spatially averaged in the white rectangle in Fig.~\ref{fig:tem}) at the start time of the event-associated type III radio burst (19:50\,UT) and three other times preceding the type III burst. The EM-weighted log temperatures in K are shown for these four times.
\label{fig:ftt}}
\end{figure}

Typical $^3$He-rich events show an enhancement pattern that increases approximately as a power law with $Q/M$ (e.g., Fig.~9 in \citet{2004ApJ...606..555M}). Perhaps these S-peaked events have more wave activity but less reconnection since reconnection produces power laws \citep{2009ApJ...700L..16D}. Models of $^3$He-rich SEP heating and acceleration by plasma waves \citep{1978ApJ...224.1048F,1992ApJ...391L.105T} have been previously used to discuss temperature and elemental abundances \citep[e.g.,][]{1997ApJ...477..940R,1980ApJ...239.1070M,2023ApJ...957..112M}. We estimate heavy-ion abundances, applying for the first time the EM-derived temperatures for the given solar source, in models of \citet{1978ApJ...224.1048F} and \citet{1997ApJ...477..940R}.

In the \citet{1978ApJ...224.1048F} model, $^3$He enrichment is achieved through heating by electrostatic ion cyclotron waves excited in the frequency range near the $^3$He gyrofrequency (i.e., between $^4$He$^{+2}$ and H cyclotron frequencies). $^4$He$^{+2}$ and H are also heated, but their thermal speeds are significantly lower than $^3$He$^{+2}$. Heavy ions, partially stripped of their electrons, have a second harmonic of their gyrofrequency in the same frequency range as $^3$He and can be heated too. For example, $^{24}$Mg with eight electrons removed has the second harmonic of its gyrofrequency equal to the $^3$He gyrofrequency. The wave excitation requires relatively high He/H ($\gtrsim$0.2), a low ($<$10$^{-3}$) plasma $\beta$, and a ratio of electron to ion temperature $T_{\rm e}/T_{\rm i}<$10. Heated ions are then injected into the (unspecified) flare acceleration process. Based on the model of \citet{1978ApJ...224.1048F}, Fig.~\ref{fig:ftt}a shows the fractions of the ionization states of selected ion species that fall within the limits given by formula (6) in \citet{1980ApJ...239.1070M} as a function of temperature, assuming the most recent equilibrium charge states from the CHIANTI atomic database \citep{1997AAS..125..149D,2021ApJ...909...38D} and the relative isotopic abundances given by \citet{2003ApJ...591.1220L}. Formula (6), repeated below from \citet{1980ApJ...239.1070M}, gives the frequency range for which species i will be resonantly heated.

\begin{equation}
1.05<2\Omega_{\rm i}/\Omega_{\rm He}<1.19+0.13(m_{\rm He}/m_{\rm i})^{1/2}
\end{equation}
where $\Omega_{\rm i}$ is gyrofrequency, $m_{\rm i}$, and $m_{\rm He}$ are the masses of the ion and $^4$He, respectively. It shows lower and upper bound frequencies for ion second harmonics in units of $^4$He gyrofrequency. The formula was derived from the minimum critical electron speed to excite electrostatic ion cyclotron waves. A lower bound of 1.05 ensures $T_{\rm e}/T_{\rm i}<$10. The upper bound was obtained for $T_{\rm e}/T_{\rm i}=$5, the value where $^{3}$He is effectively heated, as discussed by \citet{1978ApJ...224.1048F}. Note that $^{12}$C does not have gyrofrequencies in the range of frequencies for preferential heating given by (1). Only minor isotope $^{13}$C (and with $Q=+$4) contributes, which has an abundance of 1\% of $^{12}$C \citep{2003ApJ...591.1220L}.

\citet{1992ApJ...391L.105T} and \citet{1997ApJ...477..940R} proposed a model of direct acceleration (where no heating is required) of $^3$He$^{+2}$ by electromagnetic cyclotron waves that propagate with frequency between the H and $^4$He$^{+2}$ gyrofrequencies. The abundant species H and $^4$He$^{+2}$ are not accelerated because of cyclotron damping of the waves. Heavy species with $Q/M<$0.5 are accelerated at the second or higher harmonics of their cyclotron frequency. Contrary to the model of \citet{1978ApJ...224.1048F}, here a wide range of gyrofrequencies is available for acceleration. Figure~\ref{fig:ftt}b shows the fractions of the ionization states of selected ion species with $Q/M<$0.49, i.e., those that fall below the damped wave region. \citet{1997ApJ...477..940R} noted that the heavier ions are more likely to have a charge-to-mass ratio of less than 0.50 (see also Fig.~7 in \citet{2023ApJ...957..112M}) and, thus, are more likely to be accelerated. Furthermore, the authors pointed out that thermal heavy ions can be removed from resonance by Coulomb collisions whose rate is equivalent to $Q^2/M$.

Figure~\ref{fig:ftt}c plots the spatially averaged EM curves in a rectangle surrounding the solar source at different times. The EM curve peaks at log~$T$ (K) $=$6.1 ($\sim$1.3\,MK) at 19:47\,UT, log~$T$ (K) $=$ 6.2 ($\sim$1.6\,MK) at 19:48\,UT (time of mini filament eruption), and log~$T$ (K) $=$ 6.3 ($\sim$2\,MK) at 19:49/19:50\,UT (time of type III burst). The source region is multithermal, as suggested by the broad distribution of EM curves. The broadening is most obvious around type III radio burst time. For example, at 19:49\,UT, there is 0.3\,MK plasma in the source with a factor of $\sim$3 lower EM than the peak (2\,MK) temperature. At the same time, there is also a $\sim$20\,MK component with EM a factor of $>$10 lower than the peak (2\,MK) temperature.

Using EM-determined temperature, we estimate the fractional abundances of ions for preferential heating or acceleration in 2023 October 24 event. Dominant temperature (i.e., temperature with maximum EM) would not be a good choice as this neglects other temperatures at the source. We may select fractional abundance at EM-weighted temperatures. However, a single temperature, even EM-weighted, does not reflect fractional abundance profiles with temperature. Fractional abundances at various temperatures should be considered as plasma in the source is multithermal. Therefore, we multiply EM (normalized to the area under the EM curve) with fractional abundance curves shown in Fig.~\ref{fig:ftt}a,b. The integral of this multiplication over the temperature range yields the average fractional abundance. For EM at 19:49\,UT and fractional abundances in Fig.~\ref{fig:ftt}a, the average fractional abundances in percent are 0.24 (C), 0.44 (N), 2.20 (O), 3.36 (Ne), 5.46 (Mg), 12.1 (Si), 17.0 (S), 23.2 (Ca), and 18.5 (Fe). The values obtained from Fig.~\ref{fig:ftt}b are 40.1 (C), 66.4 (N), 85.2 (O), 95.3 (Ne), 97.1 (Mg), 97.9 (Si), 99.2 (S), and 100 (Ca--Fe).

\begin{figure}
\plotone{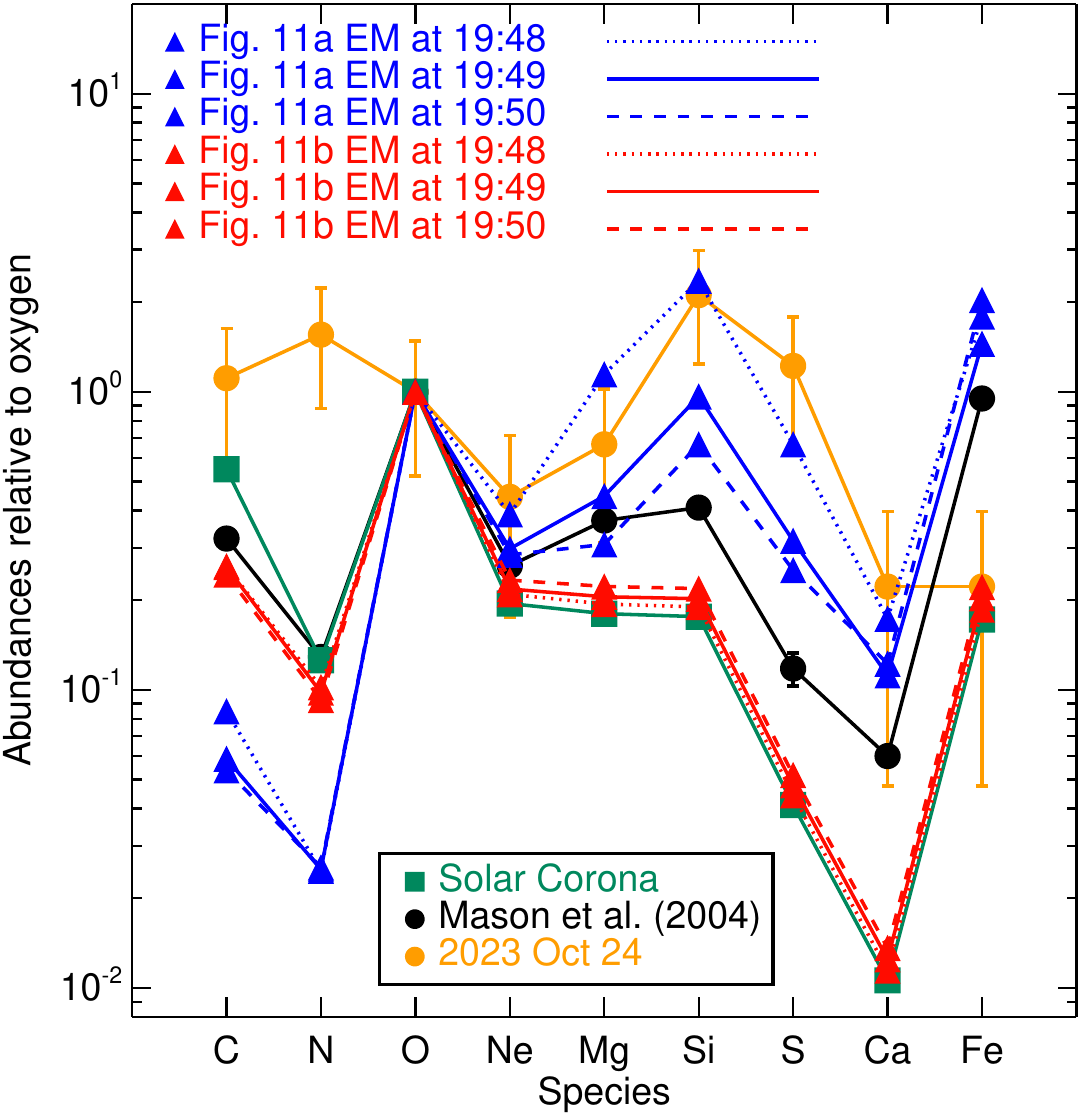}
\caption{Estimated abundances from fractions in models in Fig.~\ref{fig:ftt}, using EM at 19:48 (dotted lines), 19:49 (solid lines), 19:50\,UT (dashed lines), and solar corona abundances. Overplotted are 0.193\,MeV\,nucleon$^{-1}$ 2023 October 24 event, 0.386\,MeV\,nucleon$^{-1}$ $^3$He-rich SEP \citep{2004ApJ...606..555M}, and solar corona \citep{2012ApJ...755...33S} abundances.
\label{fig:det}}
\end{figure}

We apply these average fractional abundances to the recommended solar corona composition reported by \citet{2012ApJ...755...33S} to estimate the abundances of heated or accelerated populations. We normalized the average fractional abundances to oxygen and multiplied them with solar corona abundances, which were also normalized to oxygen. Figure~\ref{fig:det} shows the derived abundances at 19:48, 19:49, and 19:50\,UT from fractions in the two models in Fig.~\ref{fig:ftt}a,b and, for the comparison, the 0.193\,MeV\,nucleon$^{-1}$ 2023 October 24 event abundances. The comparison is adequate for low-energy ions where charge states depend only on temperature \citep[e.g.,][]{2000AA...357..716K} and not on charge stripping effects causing the energy dependence of heavy-ion charge states observed in $^3$He-rich events at higher energies, $\sim$0.2--1\,MeV\,nucleon$^{-1}$ \citep[e.g.,][]{2007SSRv..130..273K,2008ApJ...687..623D}. Overplotted in Fig.~\ref{fig:det} are 0.386\,MeV\,nucleon$^{-1}$ $^3$He-rich SEP \citep{2004ApJ...606..555M} and solar corona values. The average fractional abundance of N cannot be correctly determined from the fractions in Fig.~\ref{fig:ftt}a as the N curve is located mostly below the minimum EM-derived temperature. Thus, the abundance of N could be treated as underestimated. For the same reason, the abundance of O is also partially affected. 

\subsubsection{Fisk model} \label{subsec:fis}

Figure~\ref{fig:det} shows that the pattern of heavy-ion (Ne--Ca) abundances estimated from fractions in the \citet{1978ApJ...224.1048F} model (Fig.\ref{fig:ftt}a) is consistent with measurements with a peak at Si. Within uncertainties, the estimated abundances at the mini filament eruption (19:48\,UT) agree with the observed abundances. Thus, the specific EM distribution of temperatures could be one cause of the unusual heavy ion abundance enhancement pattern. Around the time of the type III radio burst (19:49--19:50\,UT), the pattern remains preserved, but the estimated abundances decrease, and Ne and Mg approach typical $^3$He-rich SEP values. It may raise the question of whether enhanced composition was produced early in the solar event, before type III burst onset. The Fe/O in the examined event is consistent with the solar corona value and well below the typical $^3$He-rich SEP and the estimated ratio. It may suggest that Fe is not preferentially heated, though the average fractional abundance of Fe is the second highest among other species. The high abundance of C would be hardly reproduced with its maximum fractional abundance of $\sim$1\% (see Fig.~\ref{fig:ftt}a) and therefore remains puzzling. 

\subsubsection{Roth and Temerin model} \label{subsec:tem}

Figure~\ref{fig:det} shows that the red curves, the estimated relative abundances from fractions in a model of \citet{1997ApJ...477..940R} (Fig.~\ref{fig:ftt}b), all lie almost on top of each other. Except for C (and partially N), the red lines lie close to the solar corona values. It is because the average abundance fractions of heavy ion species are close or equal to each other (i.e., all ions participate equally), placing the abundances relative to O at unity. Thus, the temperature (or its EM distribution) in the examined source could not produce the required abundances in the \citet{1997ApJ...477..940R} model. However, a hot source, where the average fractional abundances of heavier ions decrease, would create an enhancement in the relative abundance of Ne--Fe, similar to those obtained from the \citet{1978ApJ...224.1048F} model. 

The fractional abundance of C is below or equal to O for log $T$ (K) $\lesssim$7.1 ($\sim$12.6\,MK; see Fig.\ref{fig:ftt}b), implying that the relative fractional abundance of C could be as high as unity (at a cooler source). However, above that temperature, the fractional abundance of C relative to O is greater than one but less than $\sim$3 (not visible in Fig.~\ref{fig:ftt}b). It can easily cause a great enhancement of C above typical $^3$He-rich SEP values. Below log $T$ (K) $\sim$5.9 ($\sim$0.8\,MK; see Fig.~\ref{fig:ftt}b), where the fractional abundance of C is one, only faint relative enhancement of C above a typical $^3$He-rich SEP value can be produced. For N, the threshold temperature is log $T$ (K) $\sim$7.4 ($\sim$25\,MK; not shown), where the fractional abundance of N relative to O is slightly above one, but $\lesssim$1.5. It can hardly lead to an enhancement of N above the typical $^3$He-rich SEP value from a baseline coronal composition. Below log $T$ (K) $\sim$6.1 ($\sim$1.3\,MK; see Fig.~\ref{fig:ftt}b), where the fractional abundance of N is one, the relative abundance of N would not exceed the typical $^3$He-rich SEP value.

The above discussion suggests that high temperatures are required to produce abundance enhancement in the \citet{1997ApJ...477..940R} model. For example, ions could be accelerated only at a certain altitude at the source where temperature is high, though spatially averaged EM is low. Using spatially averaged EM is a simplification. Maybe ions are accelerated in spots within the source region, where EM peaks at high temperatures. In both these cases, the \citet{1997ApJ...477..940R} model could be adequate for the examined event. Sites with a hot temperature (e.g., with EM peaking at $\sim$10\,MK) would produce enhancement of heavier species (Ne--Fe) and C. This idea is consistent with \citet{2023ApJ...957..112M} suggestion that heavy-ion enhancement is mainly due to O depletion at high temperatures in the \citet{1997ApJ...477..940R} model.

The emphasis on temperature affecting the ion $Q/M$, followed by many earlier works, may not be central in the heating and acceleration of $^3$He-rich SEPs. In this case, it might be that the puzzling enhancement pattern of heavy ions does not require a special temperature or temperature range. \citet{2018ApJ...862....7M} suggested a mechanism that is sensitive to temperature, but heavy ion enhancement is due to their lower ionization losses after the passage of a small amount of material. The combined stochastic acceleration with Coulomb losses has been suggested for the enrichment of heavy and ultra-heavy ions by \citet{2020ApJ...888...48K}. The model by \citet{2006ApJ...636..462L}, developed for $^3$He and $^4$He, does not require any special temperature considerations of the type called for by \citet{1978ApJ...224.1048F} or \citet{1997ApJ...477..940R}. 

The spectral forms in the \citet{2006ApJ...636..462L} model produce $^3$He/$^4$He ratios that vary greatly with energy, which may explain the lack of correlation of Fe and $^3$He enrichments \citep{1986ApJ...303..849M,1994ApJS...90..649R}, which are not addressed by the \citet{1978ApJ...224.1048F} or \citet{1997ApJ...477..940R} models \citep[see][]{1989ApJ...343..511W}. In the region of high energy spectral rollover, models such as those by \citet{2006ApJ...636..462L} can in principle produce extreme $^3$He/$^4$He ratios such as shown in Fig.~\ref{fig:his}

\section{Conclusion} \label{subsec:con}

Below we summarise key observational features of the $^3$He-rich SEP event measured at 0.47\,au by Solar Orbiter on 2023 October 24.
\begin{itemize}
 \item{Odd ion enhancement pattern, not ordered by mass, with high abundances of C, and particularly N, Si, and S, and $^4$He and Fe depleted.}
 \item{Energy spectra of all heavy ions except $^3$He measured below $\sim$0.5\,MeV\,nucleon$^{-1}$.}
\item{$^3$He/$^4$He increasing with energy, attaining the extreme value not observed in previous events.}
\item{Solar source with supergranulation scale size, characterized by a weak photospheric field at the edge of a coronal hole, where a mini filament eruption triggered a faint straight jet.}
\end{itemize}

The presented results suggest that the EM temperature distribution at the solar source could be one factor that affects the selection of ion species for preferential heating and acceleration. A simplified concept of EM spatially averaged at the source reproduces the observed abundance pattern for the 2023 October 24 event with the \citet{1978ApJ...224.1048F} model for Ne--Ca, but C, N, and Fe do not fit into the observed abundance pattern. For the \citet{1997ApJ...477..940R} model, acceleration in sites with high temperatures would be required. The enhancement of N remains puzzling in this model.

The addition of similar $^3$He-rich SEP events with abundances not ordered by mass indicates that cool mini filaments may be a common feature of such events. However, it is unclear whether the cool material of mini filaments plays a role, as it would not be needed for the \citet{1997ApJ...477..940R} model. The energy achieved by heavy ions seems to be influenced by the size and shape of jets in these events. Finally, strong photospheric fields at the source of these events are not needed for extreme $^3$He enrichments.

$^3$He-rich SEP events where heavy ion enrichments are not ordered by mass are rare. Probably they are too small at one au to be measured above $\sim$0.5--1\,MeV\,nucleon$^{-1}$. Thus, Solar Orbiter at closer distances to $^3$He-rich solar sources enables us to detect such events at higher energies and even those weak events that were missed at one au. Understanding these events is necessary for a more complete picture of the ion heating/acceleration of $^3$He-rich SEPs. 

\begin{acknowledgments}
Radoslav Bu\v{c}\'ik acknowledges support by NASA grants 80NSSC21K1316 and 80NSSC22K0757. Sargam M. Mulay acknowledges support from the UK Research and Innovation’s Science and Technology Facilities Council under grant award numbers ST/T000422/1 and ST/X000990/1. Solar Orbiter is a mission of international cooperation between ESA and NASA, operated by ESA. The Suprathermal Ion Spectrograph (SIS) is a European facility instrument funded by ESA under contract number SOL.ASTR.CON.00004 with CAU. We thank ESA and NASA for their support of the Solar Orbiter and other missions whose data were used in this paper. Solar Orbiter post-launch work at JHU/APL and the Southwest Research Institute is supported by NASA contract NNN06AA01C and at CAU by German Space Agency (DLR) grant \# 50OT2002. The UAH team acknowledges the financial support by the Spanish Ministerio de Ciencia, Innovacion y Universidades MCIU/AEI Project PID2019- 104863RBI00/AEI/10.13039/501100011033. CHIANTI is a collaborative project involving George Mason University, the University of Michigan (USA), the University of Cambridge (UK), and NASA Goddard Space Flight Center (USA). This study benefits from discussions within the International Space Science Institute (ISSI) Team ID 425 Origins of $^3$He-rich SEPs that concluded in 2022.

\end{acknowledgments}

\bibliography{ads}{}
\bibliographystyle{aasjournal}



\end{document}